\documentclass[twocolumn,showpacs]{revtex4-2}

\usepackage{amsmath}
\usepackage{amssymb}
\usepackage{bm}
\usepackage{epsfig}
\usepackage{graphicx}
\usepackage{color}
\usepackage[colorlinks=true,citecolor=blue,urlcolor=black]{hyperref}
\usepackage[english]{babel}

\begin{document}

\title{Coherent spin dynamics of excitons in strained monolayer semiconductors}

\author{M. M. Glazov}

\affiliation{Ioffe Institute, 194021 Saint Petersburg, Russia}


\begin{abstract}
We develop a model of the {coherent} exciton spin-valley dynamics in two-dimensional transition metal dichalcogenides under elastic strain. The strain  splits the exciton radiative doublet in linearly polarized states. Consequently, it induces an effective magnetic field acting on the exciton pseudospin and causes its precession. As a result, {under circularly polarized excitation, the} circular polarization of excitons oscillates with time{, also a} time-oscillating linear polarization appears. We study competition of the strain-induced effective magnetic field with the field caused by the exciton longitudinal-transverse splitting. We uncover different regimes of  coherent spin dynamics of two-dimensional excitons{. In particular, we show} that for sufficiently large strain-induced and longitudinal-transverse splittings two frequencies related to these splittings appear in the exciton circular polarization beats.
\end{abstract}

\maketitle

\section{Introduction}

Back in 1958 E.I. Rashba predicted the so-called resonant, also termed as annihilation, interaction between electron and hole in Wannier-Mott excitons~\cite{rashba_exchange}. This work helped to resolve contradictions between previous works on the exciton fine structure within the effective mass approximation~\cite{DRESSELHAUS195614} and general theory of polaritons in semiconductors~\cite{pekar58,PhysRev.112.1555,pekar59}. Paper~\cite{rashba_exchange}  demonstrated the existence of the long-range exchange interaction between the electron and hole in semiconductors and  laid a firm basis for a consistent theory of the exchange interaction in excitons and their energy spectrum fine structure developed in the later works~\cite{BP_exch71,denisovmakarov,zhilich72:eng}. 

Nowadays the electron-hole exchange interaction is actively studied in the context of semiconductor quantum wells~\cite{maialle93,PhysRevLett.81.2586,goupalov98}, quantum dots and nanocrystals~\cite{maialle00,goupalov:electrodyn,ivchenko05a,Goupalov06,Becker:2018aa,PhysRevApplied.11.034042,Avdeev:2020ab,PhysRevB.106.125301} and two-dimensional (2D) semiconductors~\cite{glazov2014exciton,Yu:2014fk-1,PhysRevB.89.205303,PhysRevB.90.161302,PSSB:PSSB201552211,prazdnichnykh2020control,Dyksik2021}. The growing family of 2D semiconductors includes transition-metal dichalcogenide monolayers (TMDC MLs) which are renown for their outstanding optical, excitonic, electronic properties{, and remarkable spin-valley physics; these materials can be a basis for the van der Waals heterostructures providing additional degrees of control of spin and valley degrees of freedom} ~\cite{Xiao:2012cr,Geim:2013aa,Kolobov2016book,RevModPhys.90.021001}.

Electronic properties of 2D materials can be manipulated by elastic strains~\cite{PhysRevB.92.195402,PhysRevB.100.195126} that can be also applied locally~\cite{Castellanos-Gomez:2013tn} making it possible to confine~\cite{PhysRevB.97.195454,Smiri:2021wv} and steer excitons~\cite{Moon:2020vw,Hyeongwoo:vi,Bai:2020uh,Rosati:2021wt,PhysRevB.104.085405,Florian:ub}{, see also Ref.~\cite{Peng:2020aa} and references therein for review. Much less is known about the strain effect on the exciton fine structure: The splitting of the exciton radiative doublet formed of otherwise degenerate states emitting in $\sigma^+$ and $\sigma^-$ circular polarizations into linearly polarized states has been reported in Refs.~\cite{PhysRevB.99.155414,10.1088/2053-1583/ac7c21}, but a detailed systematic study of the effect has been missing until recently. In our  work~\cite{PhysRevB.106.125303} we have derived the effective Hamiltonian of the exciton radiative doublet in the presence of elastic strain and demonstrated both experimentally and theoretically the consequences of the strain on the exciton fine structure properties:} The strain modifies optical selection rules and induces, via the electron-hole exchange interaction, a fine structure splitting of the exciton radiative doublet. Interestingly, in 2D TMDCs the strain also results in the spin-dependent wavevector linear terms in the effective Hamiltonians of electrons, holes, and excitons~\cite{PhysRevB.106.125303}, similar to  Rashba spin-orbit terms in conventional semiconductors and semiconductor nanosystems~\cite{rashbasheka,rashba64,bychkov84,Rashba03}.

{The strain-induced splitting of the exciton radiative doublet should also result in coherent quantum beats of the excitons if a superposition of linearly polarized eigenstates is excited by a circularly polarized light. This effect is analogous to the electron spin precession in the external magnetic field~\cite{ivchenko05a}. The  studies of coherent spin dynamics via time- and polarization-resolved photoluminescence and time-resolved pump-probe experiments provide deep insight in the fine structure, allow one to uncover splittings hidden by an inhomogeneous broadening of optical resonances, and give an access to the kinetic processes governing relaxation between the split levels~\cite{yakovlev_bayer,amand_marie}. While calculations of the exciton energy splitting in the presence of strain and induced linear polarization are sufficient to interpret optical spectroscopy experiments such as absorption/reflection or photoluminescence, the model of coherent spin dynamics of excitons in 2D TMDCs is needed to form a basis for future time-resolved experimental studies of exciton spin dynamics in strained monolayers. Such a theory is presented in this work. Furthermore, the exciton fine structure splitting is contributed both by the exciton wavevector-independent strain-induced contribution and a wavevector-dependent exciton longitudinal-transverse splitting~\cite{glazov2014exciton,Yu:2014fk-1,PhysRevB.89.205303}. An interplay of these two contributions, as we show here, results in specific features of the exciton quantum spin beats and for sufficiently large splittings allows one to disentangle these contributions in the spin dynamics.}

Here we study theoretically manifestations of the exciton fine structure in TMDC MLs, particularly, the strain induced contributions, in the coherent spin dynamics of excitons. We analyze exciton spin beats in the presence of the strain and address an interplay of exciton longitudinal-transverse splitting and strain in exciton spin precession and damping. We briefly discuss the role of the energy relaxation processes and inhomogeneities of the strain.

\section{Model}\label{sec:model}

We recall that in 2D TMDCs  bright excitons arise from optical transitions in two valleys $\bm K_+$ and $\bm K_-$ that are active in $\sigma^+$ and $\sigma^-$ circular polarizations and superpositions of these states are active in linear polarizations~\cite{Kioseoglou,Zeng:2012ys,Sallen:2012qf,Mak:2012qf,Jones:2013tg,PSSB:PSSB201552211,PhysRevLett.117.187401,app8071157,Glazov_2021}. These two states can be mapped to the spin  $1/2$ states. Thus exciton radiative doublet can be described in terms of pseudospin~\cite{ivchenko05a}. 

Effective Hamiltonian describing a fine structure of a radiative doublet of a two-dimensional exciton in the presence of strain can be recast in the form
\begin{equation}
\label{H:eff}
\mathcal H =\frac{\hbar}{2} (\hat{\bm \sigma} \cdot \bm \Omega_{\bm K}),
\end{equation}
where $\hat {\bm \sigma} = (\hat \sigma_x,\hat \sigma_y, \hat \sigma_z)$ are the pseudospin  Pauli matrices describing the exciton pseudospin: $\hat \sigma_z$ describes the circularly polarized states,  $\hat\sigma_x$ and $\hat\sigma_y$ describe the linearly polarized components in the coordinate frames rotated by $45^\circ$ with respect to each other~\cite{ivchenko05a, glazov2014exciton}, $\bm K = (K_x, K_y)$ is the two-dimensional translational motion wavevector of the exciton (the ML is in the $xy$ plane), and $\bm \Omega_{\bm K}$ is the effective magnetic field acting on the exciton pseudospin~\cite{PhysRevB.106.125303}
\begin{subequations}
\label{Omega}
\begin{align}
&\Omega_{x,\bm K} = \mathcal A(K) (K_x^2-K_y^2) + \mathcal B (u_{xx} - u_{yy}),\\
&\Omega_{y,\bm K} = 2\mathcal A(K)K_x K_y + 2\mathcal B u_{xy},\\
&\Omega_{z,\bm K}=\mathcal C[(u_{xx} - u_{yy}) K_x - 2u_{xy} K_y].\label{Zeeman}
\end{align}
\end{subequations}
Here  $u_{ij}$ ($i,j=x,y,z$) are the Cartesian components of the strain tensor~\cite{birpikus_eng}, $\mathcal A(K)$, $\mathcal B$, and $\mathcal C$ are the parameters describing the exciton fine structure. {The parameters $\mathcal B$ and $\mathcal C$ result from the strain.} The product  $\mathcal A(K)K^2$ is responsible for the longitudinal-transverse (LT) splitting of the excitonic states~\cite{glazov2014exciton,Yu:2014fk-1,PhysRevB.89.205303,PhysRevB.101.115307,prazdnichnykh2020control}, $\mathcal A\ne 0$ in the absence of strain and we disregard possible strain-induced modification of the function $\mathcal A$. Microscopically, the main contribution to the LT splitting of the two-dimensional excitons is provided by the long-range exchange interaction between the electron and hole which can be treated as a process of (virtual) recombination and generation process of the electron-hole pair~\cite{rashba_exchange,BP_exch71,zhilich72:eng,denisovmakarov,goupalov98,goupalov:electrodyn,glazov2014exciton,prazdnichnykh2020control,PhysRevB.106.125303}. Accordingly, for a suspended monolayer,
\begin{equation}
\label{alpha}
\mathcal A(K) \approx \frac{\Gamma_0}{q K},
\end{equation}
with $\Gamma_0$ being the exciton radiative decay rate and $q=\omega_0/c$ being the light wavevector at the frequency of the exciton transition $\omega_0$. For encapsulated monolayers the long-range exchange interaction is {partially} screened and $\mathcal A(K)$ becomes reduced as compared to Eq.~\eqref{alpha}, while the wavevector dependence is approximately preserved~\cite{prazdnichnykh2020control}. Note that for the states within the light cone, $K<q$, Eq.~\eqref{alpha} is inapplicable, $\mathcal A(K)$ is imaginary and the exchange interaction results in the renormalization of the exciton radiative decay rates, such states are not considered here.

The parameters $\mathcal B$ and $\mathcal C$ describe the effects of the anisotropic strain on the exciton fine structure. Particularly, $\mathcal B$ is responsible for the splitting of excitonic states into linearly polarized components along the main axes of the strain tensor $u_{ij}$. The parameter $\mathcal C$ describes an effective magnetic field arising due to the exciton propagation in the presence of strain {that appears for the excitons with finite center of mass momentum.}

{The mixed strain components $u_{xz}$, $u_{yz}$, if present, can couple dark excitonic states with the bright ones. Here we disregard these effects for the following reasons: (i) In experimental settings like in Refs.~\cite{PhysRevB.99.155414,10.1088/2053-1583/ac7c21,PhysRevB.106.125303} these mixed strain components are expected to be small compared to the in-plane ones; (ii) Due to an interplay of the exchange and spin-orbit interaction typical energy splittings between the bright and dark states are significant and exceed by far the strain-induced splittings. Here a possible exception could be MoSe$_2$ MLs with relatively small dark-bright spltting~\cite{Robert:2020tw} which require a separate study.}

 Detailed analysis of the eigenstates of the Hamiltonian~\eqref{H:eff} and microscopic mechanisms behind the parameters $\mathcal B$ and $\mathcal C$ is presented in Ref.~\cite{PhysRevB.106.125303}. Here we focus on the exciton pseudospin dynamics in the presence of strain {and interplay of the longitudinal-transverse and strain-induced contributions in the spin beats}.

The dynamics of the radiative doublet is conveniently described in the density matrix approach~\cite{ivchenko05a,kavokin05prl,PSSB:PSSB201552211,PhysRevB.106.035302} $\varrho_{\bm K} = n_{\bm K}+(\bm s_{\bm K} \cdot \bm \sigma)$. Here $n_{\bm K}$ is the average occupancy of the state with the wavevector $\bm K$ and $\bm s_{\bm K}$ is the average value of the pseudospin in this state. Generally $n_{\bm K}$ and $\bm s_{\bm K}$ obey a set of coupled kinetic equations
\begin{subequations}
\label{kinetic}
\begin{align}
\frac{\partial n_{\bm K}}{\partial t} = Q\{n, \bm s\},\label{kin:N}\\
\frac{\partial \bm s_{\bm K}}{\partial t} + \bm s_{\bm K} \times \bm \Omega_{\bm K} = \bm Q\{\bm s, n\}.\label{kin:S}
\end{align}
\end{subequations}
where $Q\{n,\bm s\}$ and $\bm Q\{\bm s,n\}$ are the collision integrals for the exciton and pseudospin distribution functions, respectively. It follows from Eqs.~\eqref{Omega} and \eqref{kinetic} that the strain for excitonic pseudospin plays a role of {effective}  magnetic field applied in the plane of the structure, thus coherent spin beats are expected alike those arising for electron spins in the presence of real magnetic field~\cite{yakovlev_bayer}.

\section{Results and discussion}\label{sec:res}

\subsection{Analysis of parameters}

Before proceeding to the numerical and analytical results, let us briefly estimate the parameters is question. The exciton LT splitting $\Omega^{LT}_{K}$ scales approximately linearly with the exciton wavevector, see Eqs.~\eqref{Omega}  and \eqref{alpha}. Correspondingly, we present
\begin{equation}
\label{Omega:LT}
\Omega^{LT}_{K} = \mathcal A(K) K^2 = V_{LT} K.
\end{equation}
The parameter $V_{LT}$ can be associated with the effective velocity arising due to the $K$-linear LT splitting. For suspended monolayer $V_{LT} = {\Gamma_0}/{q}$, see Eq.~\eqref{alpha}. Depending on the system parameters $V_{LT} = 10^6\ldots 10^7$~cm/s.

Another parameter that has a dimension of velocity is the coefficient $\mathcal C$ at the strain-induced $\bm K$-linear Zeeman effect, Eq.~\eqref{Zeeman}. According to Ref.~\cite{PhysRevB.106.125303} the parameter $\mathcal C \sim \gamma_3/\hbar$ where $\gamma_3$ {is related to the interband momentum matrix element in the $\bm k\cdot \bm p$-model}. For $\gamma_3 \approx 3$~eV$\cdot$\AA~\cite{2053-1583-2-2-022001,PhysRevB.95.155406} we have $\mathcal C \approx 10^7$~cm/s. Since $\Omega_{z,\bm K}$ in Eq.~\eqref{Omega} is determined by the product of $\mathcal C$ and strain components, the effective strain-induced velocity can be introduced as $V_{strain} = \mathcal C (u_{xx} - u_{yy})$. For reasonable strain values which do not exceed several percent, $V_{strain} \ll V_{LT}$. Thus, in most cases, the effect of the strain-induced $\bm K$-linear terms are negligible for excitons as compared to the LT splitting. 

Next, let us compare the LT and strain-induced wavevector independent splittings. The former depends on the exciton center of momentum $K$ that, in turn, depends on the excitation conditions and temperature. It can vary, in typical conditions, from fractions of meV to tens of meV in atomically thin crystals for reasonable $K\sim 10^6$~cm$^{-1}$. From analytical estimates and experimental data reported in Ref.~\cite{PhysRevB.106.125303} for {reasonable values of uniaxial strain} the induced splitting $\hbar \mathcal B (u_{xx} - u_{yy}) \sim 1$~meV. Thus, typically, LT and strain-induced splittings can be comparable in magnitude. {It is noteworthy that while particular parameters vary between different TMDC MLs, their orders of magnitude are similar due to similarity of the crystalline and band structure. That is why we use generic values of parameters to illustrate the basic effects in coherent spin dynamics of excitons.}

The dynamics of pseudospin is controlled not only by the magnitude of $\bm \Omega_{\bm K}$ in Eqs.~\eqref{Omega} and \eqref{kin:S}, but also by two more parameters that determine the kinetics of excitons: their mean kinetic energy $\bar E$ and scattering rate $\tau^{-1}$. Typically, three parameters that have dimension of energy: $\hbar\Omega$ with $\Omega$ being the characteristic value of $|\bm \Omega_{\bm K}|$, $\bar E$ and $\hbar/\tau$ can be comparable in magnitude for excitons in 2D TMDCs. It makes kinetic processes quite involved in such systems, see, e.g., Refs.~\cite{grimaldi05,Averkiev:2008aa,PhysRevLett.124.166802,PhysRevLett.127.076801}. Here we focus on the simplified  description of the exciton pseudospin dynamics and assume that the applicability condition of the kinetic equations~\eqref{kinetic}, $\bar E \gg \hbar/\tau$, is fulfilled. Also, for simplicity, we consider $\hbar\Omega \ll \bar E$, while the product $\Omega\tau$ can be arbitrary. Depending on whether product $\Omega\tau$ is smaller or larger than unity the pseudospin dynamics can be quite different. {We stress that we are interested in the situations where excitons are excited resonantly or quasi-resonantly and where the temperature is not too high. In this case $\bar E \ll E_g$ where $E_g$ is the band gap, and it is sufficient to use the approximation of parabolic isotropic dispersion for excitons and keep lowest in the wavevector $\bm K$ contributions to the precession frequency $\bm\Omega_{\bm K}$, Eq.~\eqref{Omega}. }

\subsection{Coherent spin dynamics of excitons}

In what follows we focus on the exciton pseudospin dynamics and take the collision integral in the form
\begin{equation}
\label{collision}
\bm Q\{\bm s, n\} = - \frac{\bm s_{\bm K} - \bar{\bm s_{\bm K}}}{\tau},
\end{equation} 
where $\tau$ is the scattering time. The exciton lifetime and exciton energy relaxation is assumed to exceed by far $\tau$ and $\Omega^{-1}$ as it typically occurs for exciton scattering by acoustic phonons~\cite{shree2018exciton,PhysRevLett.124.166802}. We also select the coordinate system where $u_{xy}=0$, i.e., the $x$ and $y$ axes are the main axes of the strain, and introduce the notation $u\equiv u_{xx} - u_{yy}$.

\begin{figure}[b]
\includegraphics[width=0.9\linewidth]{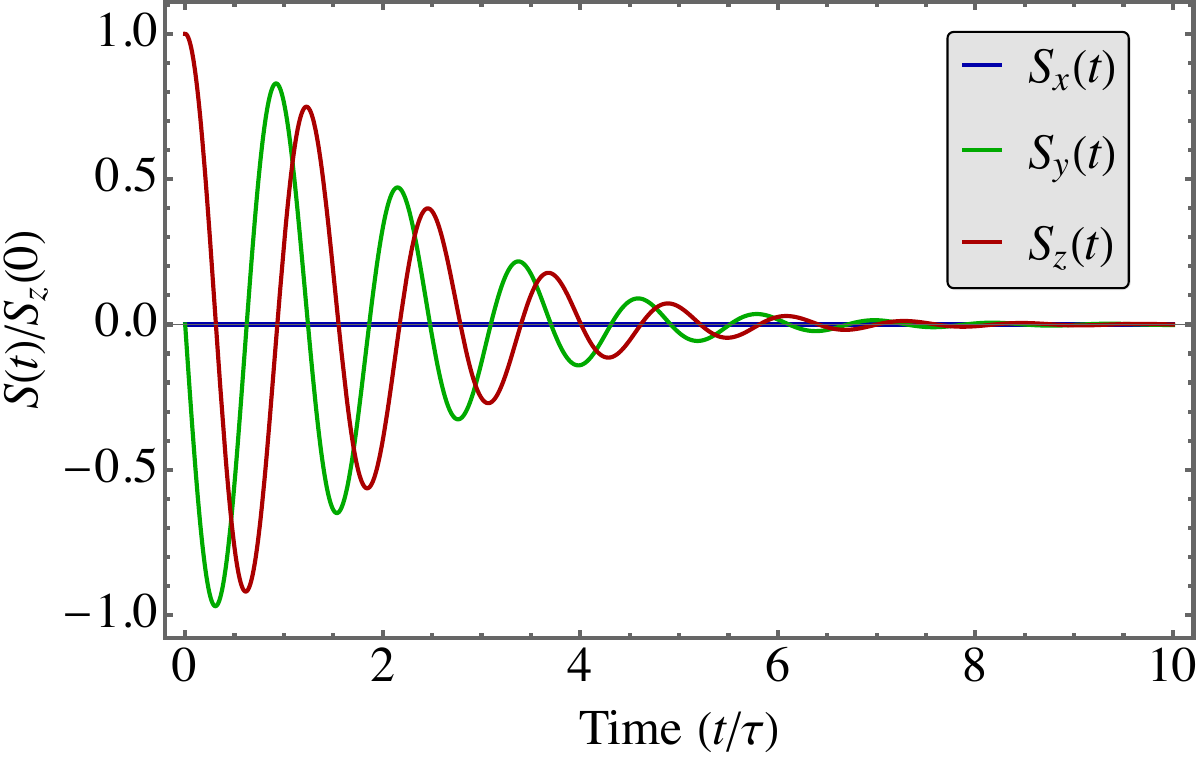}
\caption{Temporal dynamics of exciton pseudospin components $S_\alpha = \sum_{\bm K} s_{\alpha,\bm K}$ ($\alpha=x,y,z$) after pulsed circularly polarized excitation calculated by numerically solving kinetic Eq.~\eqref{kinetic}. Calculation parameters are: $\Omega_K^{LT}\tau=1$, $\mathcal B u\tau=5$, and $\mathcal C=0$.}\label{fig:1}
\end{figure}

Figure~\ref{fig:1} shows the exciton pseudospin dynamics {for the uniaxial strain} calculated numerically in the case of the {resonant} circularly polarized excitation by a short pulse: We solve the kinetic equation~\eqref{kinetic} numerically by decomposing the $\bm s_{\bm K}$ in the series of the angular harmonics with the initial condition $s_{x,\bm K} = s_{y,\bm K}\equiv 0$, $s_{z,\bm K} \propto \delta(E_K - E_0)$ [cf. Ref.~\cite{glazov2007}]. Here $E_K$ is the exciton dispersion, $E_0$ is the excitation energy and inelastic processes are neglected, thus, the energy dependence of the pseudospin distribution function is preserved. The main features of the exciton pseudospin dynamics in strained monolayers are clearly seen from Fig.~\ref{fig:1}: Pronounced  beats are present in the circular polarization $S_z$ and the conversion between the circular and linear polarization in the diagonal axes, $S_y$, is observed similarly to the polarization conversion in conventional semiconductor nanosystems~\cite{PhysRevB.56.13405,PhysRevLett.96.027402,kavokin05prl,kowalik07}. Linear polarization in the $(xy)$ axes, $S_x$, does not appear in this model, but it may arise due to  redistribution of the excitons between the fine structure split levels in the presence of inelastic processes and because of the strain-induced modification of the optical selection rules~\cite{PhysRevB.106.125303}, see also Ref.~\cite{PhysRevB.105.L241406}. {The band mixing at high exciton momenta (away from the $\bm K_\pm$ points of the Brillouin zone for electrons and holes) also results in reduction of degree of exciton optical orientation, see, e.g., Ref.~\cite{Glazov_2021}. These factors can be taken into account in our model by adjustment of initial conditions. Our model can be also applied to anisotropic  2D semiconductors where two close-in-energy excitonic states active in orthogonal linear polarization are present.}

\begin{figure}[t]
\includegraphics[width=0.9\linewidth]{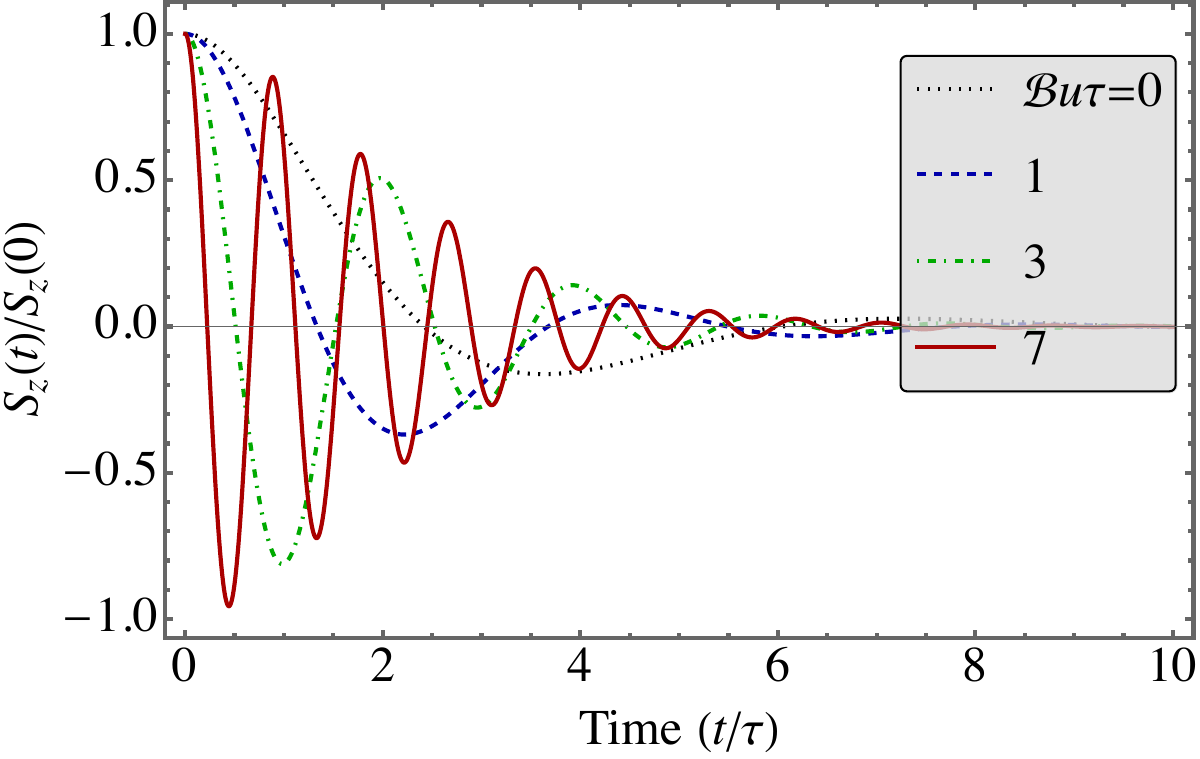}
\caption{Temporal dynamics of  $z$ component of exciton pseudospin after pulsed circularly polarized excitation calculated by numerically solving kinetic Eq.~\eqref{kinetic} for different values of elastic strain (shown in the legend). Other parameters are the same as in Fig.~\ref{fig:1}.}\label{fig:2}
\end{figure}

In Fig.~\ref{fig:2} the evolution of the exciton circular polarization dynamics as a function of elastic strain is shown. At $\mathcal Bu\to 0$ the effect of strain on the polarization beats is negligible and the dynamics is solely controlled by the long-range exchange interaction. In this case it can be described by the analytical formula (see Ref.~\cite{prazdnichnykh2020control} and references therein):
\begin{equation}
\label{u0:sz}
\frac{s_{z,K}(t)}{s_{z,K}(0)} = e^{-t/2\tau}\left(\cosh{\frac{wt}{2\tau}} + w^{-1} \sinh{\frac{wt}{2\tau}} \right),
\end{equation}
where $w=\sqrt{{(\Omega_K^{LT}\tau)}^2-1/4}$. Particularly, for $\Omega_K^{LT}\tau \gg 1$ the beats due to the exciton LT splitting are observed {and their damping is determined by the scattering time (decay rate is $2\tau$). It is because a single scattering of exciton results in significant variation of $\bm \Omega_{\bm K}$ due to the change of the exciton wave vector, thus on the time scale $\sim \tau$ the coherence in spin dynamics breaks}. At $\mathcal Bu \gg \Omega_K^{LT}$ the beats occur at the strain-related frequency, $\mathcal Bu$. In the intermediate regime the situation is more involved, see below.

\begin{figure}[t]
\includegraphics[width=0.9\linewidth]{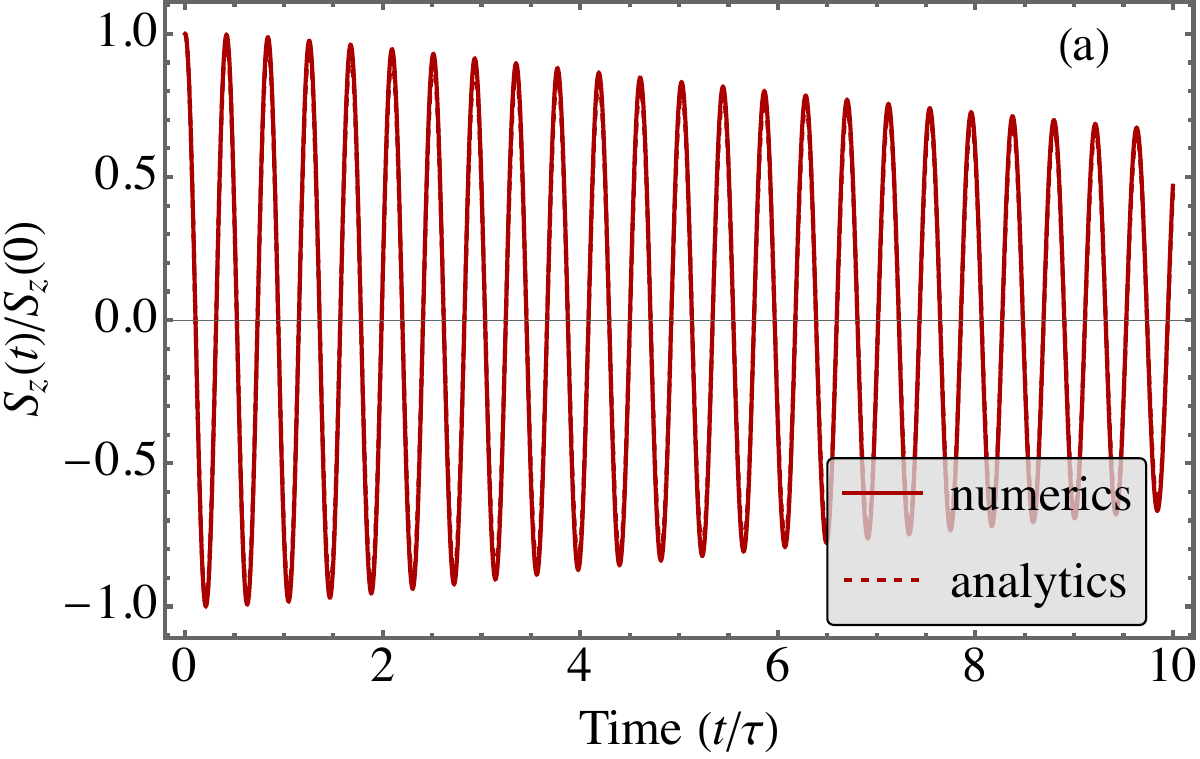}
\includegraphics[width=0.9\linewidth]{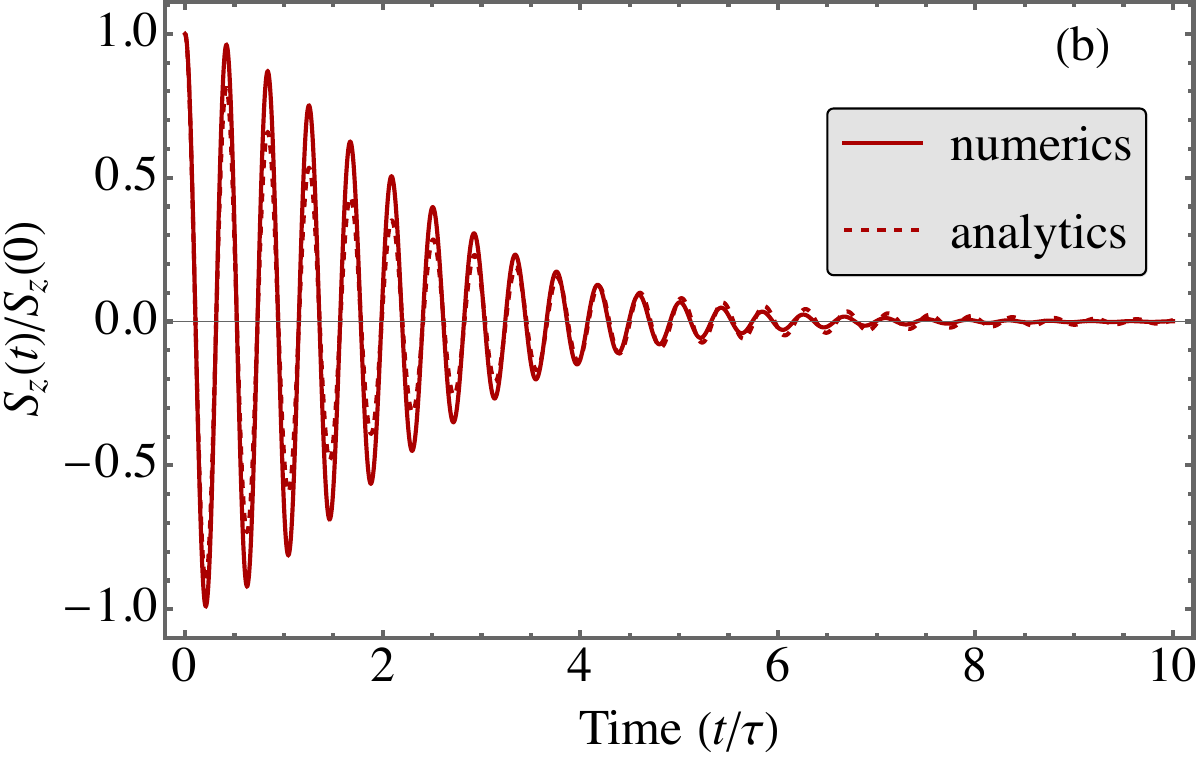}
\includegraphics[width=0.9\linewidth]{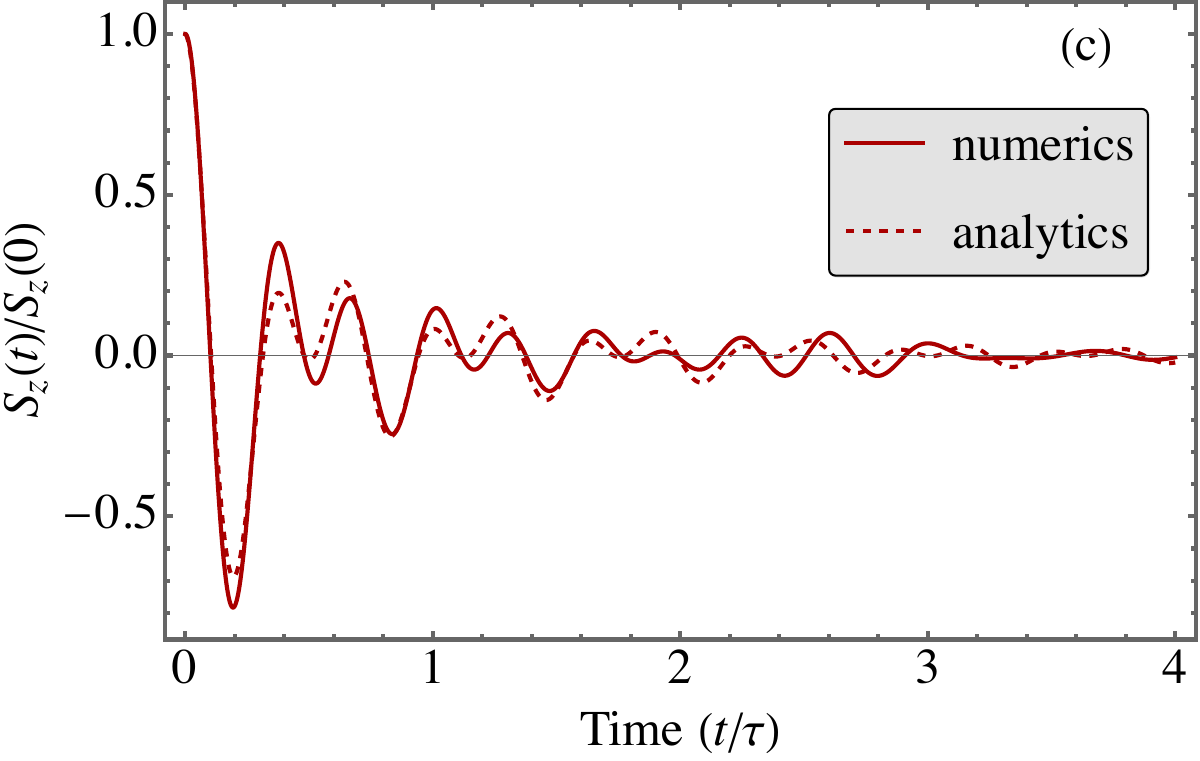}
\caption{Temporal dynamics of  $z$ component of exciton pseudospin after pulsed circularly polarized excitation calculated by solving kinetic Eq.~\eqref{kinetic} at $\mathcal Bu\tau=15$ (solid lines). (a) $\Omega_K^{LT}\tau=0.3$, (b) $\Omega_K^{LT}\tau=1$, (c) $\Omega_K^{LT}\tau=5$ Fig.~\ref{fig:1}. Solid curves present numerical results, dotted curves present analytical expressions: Eq.~\eqref{collision:dominated} [panel (a)], Eq.~\eqref{approx:intermediate} [panel (b)], and Eq.~\eqref{approx:high} [panel (c)].}\label{fig:3}
\end{figure}

The role of the LT splitting in the strain-induced exciton pseudospin dynamics is illustrated in Fig.~\ref{fig:3}. Panel (a) shows the results for the circular polarization beats in the  collision dominated regime, $\Omega_K^{LT} \tau=0.3$. In this case the exciton spin dynamics is described by analytical expressions similar to those for electron spin beats in magnetic field at anisotropic spin relaxation~\cite{Kalevich:1997aa,glazov08a}:
\begin{equation}
\label{collision:dominated}
\frac{s_{z,K}(t)}{s_{z,K}(0)} = \left[\cos{\bar\Omega t} - \frac{\Gamma_{zz} - \Gamma_{yy}}{2\bar\Omega}\sin{\bar\Omega t}\right]e^{-\bar\Gamma t},
\end{equation}
where 
\begin{equation}
\label{bars:collision}
\bar \Omega = \sqrt{(\mathcal Bu)^2-(\Gamma_{zz} - \Gamma_{yy})^2/2}, \quad 
\bar\Gamma = (\Gamma_{zz} + \Gamma_{yy})/2,
\end{equation}
 and the components of the pseudospin relaxation rates tensor are given by [cf. Ref.~\cite{glazov04}]
\begin{subequations}
\label{Gamma:collision}
\begin{align}
\Gamma_{zz} = \frac{(\Omega_K^{LT})^2\tau}{2}\left(1+ \frac{1}{1+(\mathcal Bu)^2\tau^2}\right),\\
\Gamma_{yy} =  \frac{(\Omega_K^{LT})^2\tau}{2} + \frac{(\mathcal CuK)^2\tau}{2[1+(\mathcal Bu)^2\tau^2]},\\
\Gamma_{xx} = \frac{(\Omega_K^{LT})^2\tau+ (\mathcal CuK)^2\tau}{2[1+(\mathcal Bu)^2\tau^2]}.
\end{align}
\end{subequations}
In this regime, generally, the pronounced spin beats with the frequency $\mathcal Bu$ are observed and their damping is described by the $\bar \Gamma \ll \Omega, \tau^{-1}$ in Eq.~\eqref{bars:collision} similarly to the Dyakonov-Perel spin relaxation in systems with Rashba and Dresselhaus spin-orbit splitting~\cite{dyakonov72,dyakonov86,Ivchenko73,1988JETPL..47..486I,glazov04}.  

Figure~\ref{fig:3}(b) shows an intermediate regime where $\Omega_K^{LT}\tau=1$. For $\Omega_K^{LT}\tau \sim 1$ one can use the following approximate formula to describe the spin beats 
\begin{equation}
\label{approx:intermediate}
\frac{s_{z,K}(t)}{s_{z,K}(0)} = \cos{(\Omega_{c} t)}e^{-\gamma t},
\end{equation}
that captures the main features of the pseudospin dynamics. The spin beats occur at the combination frequency $\Omega_{c} = \sqrt{(\mathcal Bu)^2+(\Omega_K^{LT})^2}$ and decay with the rate $\gamma \approx 1/(2 \tau)$.

Finally, in Fig.~\ref{fig:3}(c) the polarization dynamics in the regime where $\Omega_K^{LT} \tau = 5$ is presented. Here the pronounced beating of exciton pseudospin is seen with two distinct frequencies. A compact  expression describes the polarization dynamics can be derived in the limit where $\tau^{-1} \ll \Omega_K^{LT} \ll \mathcal Bu$:
\begin{equation}
\label{approx:high}
\frac{s_{z,K}(t)}{s_{z,K}(0)} = \cos{\left(\mathcal Bu t\right)}\mathrm J_0\left(\Omega_K^{LT}t\right)e^{-\gamma t},
\end{equation}
with $\mathrm J_0(x)$ being the Bessel function. Equation~\eqref{approx:high} in the limit of $\tau\to\infty$ can be readily obtained averaging the temporal dynamics of the exciton spin-$z$ with a given wavevector $\bm K$: $s_{z,\bm K}(t) \propto \cos{(\Omega_{\bm K} t)}$ over the directions of $\bm K$. The damping rate $\gamma \approx 1/(2\tau)$ in Eqs.~\eqref{approx:intermediate} and \eqref{approx:high} can be qualitatively understood taking into account approximately equal relative weights of the zeroth angular harmonic, which is not affected by the collision integral~\eqref{collision}, and other harmonics, which relax with the rate $\tau^{-1}$.

\subsection{Brief discussion}

We have studied above an interplay of the exciton LT splitting and strain induced splitting in coherent spin dynamics for monoenergetic quasiparticles with the same $K =|\bm K|$. Such a simple regime can be realized in the absence of inelastic processes which are indeed weak if the dominant scattering processes are related to the exciton interaction with the long-wavelength phonons and static disorder~\cite{PhysRevLett.124.166802,https://doi.org/10.48550/arxiv.2208.08501}.

Let us now briefly address the role of the energy relaxation processes. The spread of exciton energies in the course of the energy relaxation leads to a spread of the LT splittings $\Omega_K^{LT}$ and suppresses its contribution to the polarization beats. Thus, in real systems the observation of the beatings described by Eq.~\eqref{approx:high} can be hampered by the redistribution of excitons over a range of energies. 

On the other hand, the strain-induced splitting is wavevector independent. Hence, the spin beats caused by the strain are almost insensitive to the energy relaxation processes as long as the dependence of $\mathcal B$ on the exciton energy can be disregarded. The damping of these beats can be, however, affected by the energy relaxation processes to the same extent as the energy relaxation processes affect spin relaxation in conventional semiconductors. For instance, in the regime $\tau^{-1} \ll \Omega_K^{LT} \ll \mathcal Bu$, Eq.~\eqref{approx:high}, a fast initial energy relaxation (or thermalization) of excitons can be accounted for by averaging $\mathrm J_0\left(\Omega_K^{LT}t\right)$ over appropriate energy distribution of excitons. For fully thermalized excitons to the temperature $T$ we obtain 
\[
\left\langle\mathrm J_0\left(\Omega_K^{LT}t\right)\right\rangle = \exp{\left(-\frac{(\Omega_{K_T}^{LT})^2 t^2}{4} \right)}, \quad K_T = \sqrt{\frac{2 M k_B T}{\hbar^2}}.
\]
Here $M$ is the exciton translational mass {(typically $M \sim m_0$ the free electron mass in TMDC MLs)} and $K_T$ is the thermal wavevector. Correspondingly, $\Omega_{K_T}^{LT}$ gives the rate of the polarization decay in this regime. By contrast, in the regime $\Omega_{K_T}^{LT} \tau \ll 1$ Eqs.~\eqref{collision:dominated} and \eqref{bars:collision} hold but the rates $\Gamma_{ij}$ in Eq.~\eqref{Gamma:collision} should be appropriately averaged over the energy distribution of the excitons in the regime of fast energy relaxation as compared the pseudospin relaxation, $\tau_\epsilon \bar \Gamma \ll 1$, or Eq.~\eqref{collision:dominated} should be averaged over the thermal distribution in the opposite limit $\tau_\epsilon \bar\Gamma \gg 1$ with $\tau_\epsilon$ being the energy relaxation time.

We also note that in real structures the components of the strain tensor $u_{ij}$ can be random functions of coordinates due to inevitable inhomogeneities of the systems. Such spatial fluctuations of strain produce spatial fluctuations of the exciton fine structure splitting and act similarly to the random Rashba fields and spin-orbit coupling disorder in quantum wells~\cite{Sherman_RandomRashba,GlazovSherman_rev}. The strain fluctuations produce a lower bound to the exciton spin or valley relaxation rate
\begin{equation}
\gamma_{\rm min} \sim 
\begin{cases}
\mathcal B^2 \langle \delta u^2\rangle\tau_c, \quad \tau_c \ll \tau,~~\mathcal B^2 \langle \delta u^2\rangle\tau_c^2 \ll 1,\\
\mathcal B \sqrt{\langle \delta u^2\rangle}, \quad \mathcal B^2 \langle \delta u^2\rangle \gg \tau^{-2}, \tau_c^{-2}
\end{cases}
\end{equation}
Here $\langle\delta u^2\rangle$ is the mean square of the strain fluctuation, $\tau_c = l_c/(\hbar K/M)$ is the correlation time with $l_c$ being the correlation length of the strain fluctuations, and we assumed that $\langle \delta u\rangle =0$. A spatial dependence of strain and, hence, of the exciton fine structure splitting can be particularly important in moire structures formed in TMDC bilayers~\cite{Tartakovskii:2020aa}.

\section{Conclusion and outlook}

In this work we have focused on the temporal dynamics of exciton spin polarization. Under the steady-state excitation conditions a pronounced effect of the strain on the exciton optical orientation is expected. Similarly to Ref.~\cite{PhysRevB.84.073301} the exciton optical orientation can be a non-monotonous function of the strain due to a competition of the LT and strain-induced splittings of excitonic states. An interplay of the LT and strain-induced contributions can be also observed in the exciton pseudospin fluctuations, cf.~\cite{PhysRevB.90.085303,glazov_sns_pol}, and affect  valley polarization bistability and its stochastic switching~\cite{PhysRevB.106.035302}.

The exciton spin beats studied here can be observed in time-resolved experiments where the superposition of the strain-split states is created by a short circularly polarized light pulse resonant or quasi-resonant with the excitonic transition. The dynamics can be monitored by the time-resolved circular and linear polarization degree of exciton emission or via the Kerr or Faraday rotation of the polarization plane of the additional linearly polarized probe pulse, see Refs.~\cite{PhysRevB.90.161302,yakovlev_bayer,amand_marie} for details.

In conclusion, we have theoretically studied  exciton spin dynamics in two-dimensional  semiconductors based on transition metal dichalcogenides focusing on the strain effects. We have demonstrated an interplay of the exciton radiative doublet longitudinal-transverse  and the strain-induced splittings in the spin beats of excitons. The regimes where the exciton circular polarization beats are controlled by the strain have been identified. In the case where both longitudinal-transverse and strain-induced splittings of the exciton radiative doublet are sufficiently large, two frequencies in the exciton spin beats appear. The damping of the exciton spin precession has been also analyzed. Compact analytical expressions which describe numerical results are presented. A role of energy relaxation and strain fluctuations has been discussed.

\acknowledgements

This work has been supported by the RSF Project No. 19-12-00051. The author is grateful to Alexey Chernikov for valuable discussions.


%

\end{document}